\documentclass[
    ,final            
  ]
  {aipproc}

\usepackage{amsmath}
\usepackage{amssymb}

\layoutstyle{6x9}

\newcommand{\etal}{{\em et al.}}                
\newcommand{\beq}{\begin{equation}}
\newcommand{\eeq}{\end{equation}}

\begin{document}

\title{Symmetry Energy\\ as a Function of Density and Mass}


\author{Pawel Danielewicz}{
  address={National Superconducting Cyclotron Laboratory and
Department of Physics and Astronomy, Michigan State University,
East Lansing, Michigan 48824, USA}
}

\author{Jenny Lee}{
  address={National Superconducting Cyclotron Laboratory and
Department of Physics and Astronomy, Michigan State University,
East Lansing, Michigan 48824, USA}
}

\begin{abstract}
Energy in nuclear matter is, in practice, completely characterized at
different densities and asymmetries, when the density dependencies of
symmetry energy and of energy of symmetric matter are specified.
The density dependence of the symmetry energy at subnormal densities
produces mass dependence of nuclear symmetry coefficient and, thus,
can be constrained by that latter dependence.  We deduce values of the
mass dependent symmetry coefficients, by using excitation energies to
isobaric analog states.  The coefficient systematic, for intermediate and high
masses, is well described in terms of the symmetry coefficient values of
$a_a^V = (31.5$--$33.5) \, \text{MeV}$ for the volume coefficient and $a_a^S = (9$--$12) \, \text{MeV}$
for the surface coefficient.  These two further correspond to the parameter values describing density
dependence of symmetry energy, of
$L \sim 95 \, \text{MeV}$ and $K_\text{sym} \sim 25 \, \text{MeV}$.
\end{abstract}

\maketitle


\section{Introduction}

Symmetry energy is first encountered in nuclear physics in the empirical nuclear energy formula as the term describing
drop in nuclear binding with increasing neutron-proton asymmetry.  That term is quadratic in the asymmetry, expressing
the charge symmetry, symmetry of nuclear interactions under neutron-proton interchange.
In nuclear matter, the energy as a function of neutron and proton densities may be expanded in the matter asymmetry.  Knowledge of density
dependence of the expansion coefficient and of the energy of symmetric matter suffices in practice
for the determination of nuclear energy and pressure at any density and asymmetry, which is essential for
calculations of neutron star structure \cite{Lattimer:2006xb}.  The coefficients of expansion for finite nuclei and nuclear matter are connected in a nontrivial manner.

Specifically, the density dependence of the symmetry-energy expansion coefficient in nuclear matter gets tied to a
variation with mass of the symmetry-energy coefficient in the empirical energy formula, due to changing
contributions to the energy of nuclear volume and of surface where nuclear density drops.  As a consequence, it may be possible to constrain the density-dependence of symmetry energy using the mass-dependence of
the symmetry coefficient.  In the following, we deduce the symmetry energy coefficient as a function
of nuclear mass, using nuclear energies.  Moreover, we employ Skyrme-Hartree-Fock (SHF) calculations for half-infinite matter
and for spherical nuclei, combined with Hohenberg-Kohn functional theory, to constrain the density-dependence of symmetry energy.

\section{Symmetry Coefficient and Isovector Density}

Considerations \cite{Myers:1969,Danielewicz:2003dd} of the competition between nuclear surface and volume in storing neutron-proton
asymmetry leads to the following formula for nuclear energy
\beq
E(N,Z)
 =   -a_V \, A + a_S \, A^{2/3} + \frac{a_a(A)}{A} \, (N-Z)^2 + a_C \, \frac{Z^2}{A^{1/3}} + E_\text{mic} \, ,
\label{eq:enza}
\eeq
where $E_\text{mic}$ represents microscopic contributions to the energy and where the mass-dependent
symmetry coefficient $a_a(A)$ follows from
\beq
\frac{A}{a_a(A)} = \frac{A}{a_a^V} + \frac{A^{2/3}}{a_a^S} \, .
\label{eq:AaaA}
\eeq
In the above, $a_a^V$ and $a_a^S$ are the volume and surface symmetry coefficients and Eq.~(\ref{eq:AaaA}) states that the capacity of the system for asymmetry is a sum of the interior and surface capacitances.

The energy per nucleon in nuclear matter, expanded in the relative asymmetry $\eta=(\rho_n-\rho_p)/\rho$, becomes
\beq
\label{eq:EA}
\frac{E}{A} = \frac{E_0}{A} (\rho) + \frac{E_a}{A} (\rho , \eta) \simeq \frac{E_0}{A} (\rho) + S(\rho) \, \eta^2 \, .
\eeq
The coefficient $S$ is typically expanded around the normal density as
\beq
\label{eq:Sexp}
S(\rho) = a_a^V + \frac{L}{3 \rho_0} \, \left( \rho - \rho_0 \right) + \frac{K_\text{sym}}{18 \rho_0^2} \, \left( \rho - \rho_0 \right)^2 + \ldots \, .
\eeq

Considerations \cite{dan07} of the nuclear Hohenberg-Kohn (HK) energy functional \cite{hohenberg-1964}, smoothed out to make it analytic, imply that the mass-dependent symmetry coefficient may be generally represented as
\beq
\label{eq:aaaa}
\frac{A}{a_a(A)} = \frac{1}{a_a^V}
\int \text{d} {\pmb r} \rho_{a} ({\pmb r})  \, ,
\eeq
where the isovector density $\rho_a$ is defined as
\beq
\label{eq:rhoa=}
\rho_a ({\pmb r}) = \frac{ \rho_{n}({\pmb r}) - \rho_{p}({\pmb r})}{\eta_V} \, .
\eeq
Here, $\eta_V$ is asymmetry in the interior of matter.

In the limit of weak Coulomb effects, the isovector and isoscalar densities change only within the
second order in asymmetry across an isobaric chain~\cite{dan07}.  That implies that, in the continuum limit
across medium and heavy nuclei, the proton and neutron densities can be in practice described in terms of just 4 parameters, radius parameter, difusenesses for isoscalar and isovector densities and the relative displacement of isovector and isoscalar densities.  The last 2 parameters are tied to the density dependence of symmetry energy.

In the limit of short nonlocality range in the symmetry part of the HK energy functional, compared to nuclear densities, the isovector density is
\beq
\label{eq:rhoaS}
\rho_a = \frac{a_a^V}{S(\rho)} \, \rho \, .
\eeq
\begin{figure}
  \includegraphics[height=.41\textheight]{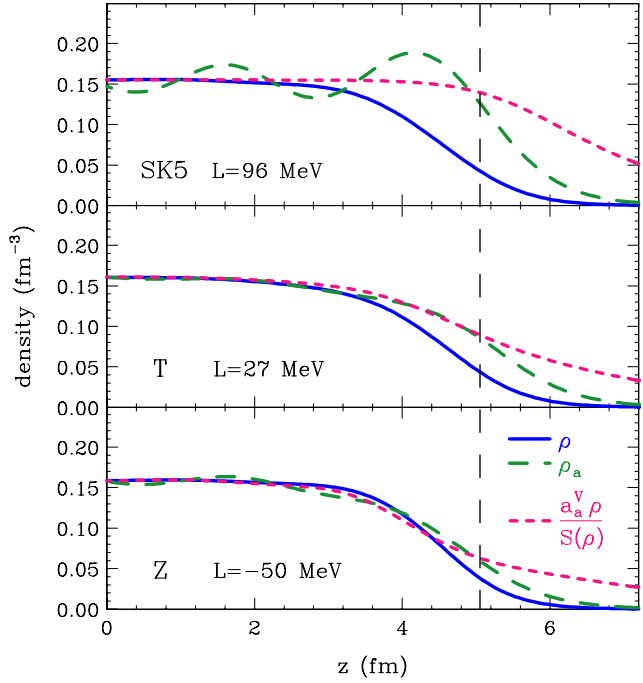}
  \caption{Isoscalar (solid curves) and isovector (dashed) density in half-infinite nuclear matter, together with a local approximation to the isovector density (short-dashed), as a function of position, for sample Skyrme interactions.}
  \label{fig:densloc}
\end{figure}
In Fig.~\ref{fig:densloc}, we show densities from our SHF calculations of half-infinite nuclear matter.  Before such calculations have been carried out Farine \etal \cite{Farine:1980}.  For $\rho \gtrsim \rho/4$, the exact isovector densities exhibit Friedel oscillations around the respective local approximations given by Eq.~(\ref{eq:rhoaS}).  This implies that differences between proton and neutron densities, and even proton densities alone, can directly express the density dependence of $S$ at $\rho > \rho/4$.  At $\rho \lesssim \rho/4$, nonlocalities in the HK energy functionals become strong as evidenced in the discrepancies between exact and approximate isovector densities.

\begin{figure}
  \includegraphics[height=.33\textheight]{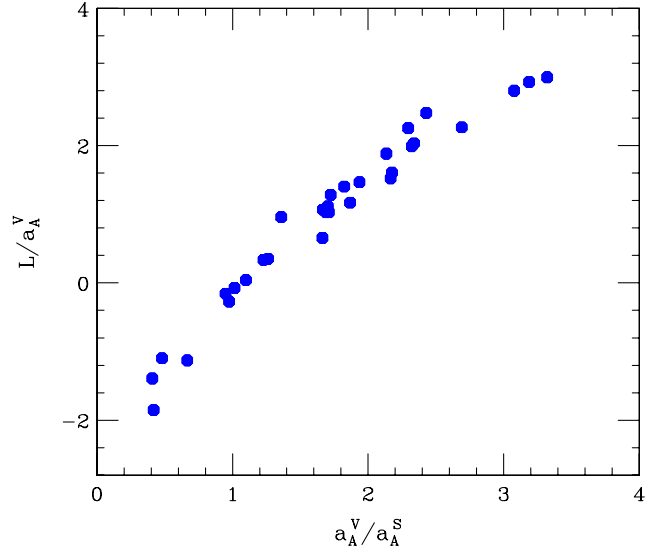}
\caption{Correlation between the slope parameter $L$ and inverse of the surface symmetry parameter~$a_a^S$, for Skyrme interactions from the compilation by Stone~\cite{PhysRevC.68.034324}.  The parameters are scaled with~$a_a^V$.}
\label{fig:lavs}
\end{figure}

From the energy of half-infinite matter and/or vector density we can determine $a_S$ and $a_a^S$ coefficients in the energy formula (\ref{eq:enza}) for the Skyrme interactions.  While the surface coefficients, $a_S \sim 19 \, \text{MeV}$, vary little between different interactions, the surface symmetry coefficients $a_a^S$ vary widely, from 9 to 75\,MeV.  The coefficients $a_a^S$ are tightly correlated to the symmetry-energy slope parameter~$L$, see Fig.~\ref{fig:lavs}.
A correlation also is found between scaled $L$ and $K_\text{sym}$.  Those correlations are likely due to the lack of other energy scale in the energy functional than~$\rho_0$.

\section{$a_a(A)$ from Isobaric Analog States}

Competition between different physics terms within an energy formula hampers the ability to learn about mass dependence of the nuclear symmetry coefficient by fitting directly a formula to the ground-state nuclear energies \cite{Danielewicz:2003dd,Danielewicz:2004et}.  However, the unwanted competition may be practically eliminated by generalizing the energy formula to lowest states of given net isospin within a nucleus \cite{Janecke:2003}, amounting to the replacement of the symmetry term in (\ref{eq:enza}):
\beq
a_a(A) \, \frac{(N-Z)^2}{A} = 4 \, a_a(A) \,  \frac{T_z^2}{A} \Rightarrow 4 \, a_a(A) \, \frac{T(T+1)}{A} \, .
\eeq
This replacement absorbs the so-called Wigner term from $E_\text{mic}$.  With the formula generalization, it becomes possible to deduce the symmetry coefficient within a single nucleus, by using excitation energies to the isobaric analog states (IAS) representing ground states of neighboring nuclei \cite{Danielewicz:2004et}, with
\beq
\Delta E = 4 \, a_a(A) \, \frac{\Delta \big( T(T+1) \big) }{A} + \Delta E_\text{mic} \, .
\eeq

\begin{figure}
  \includegraphics[height=.37\textheight]{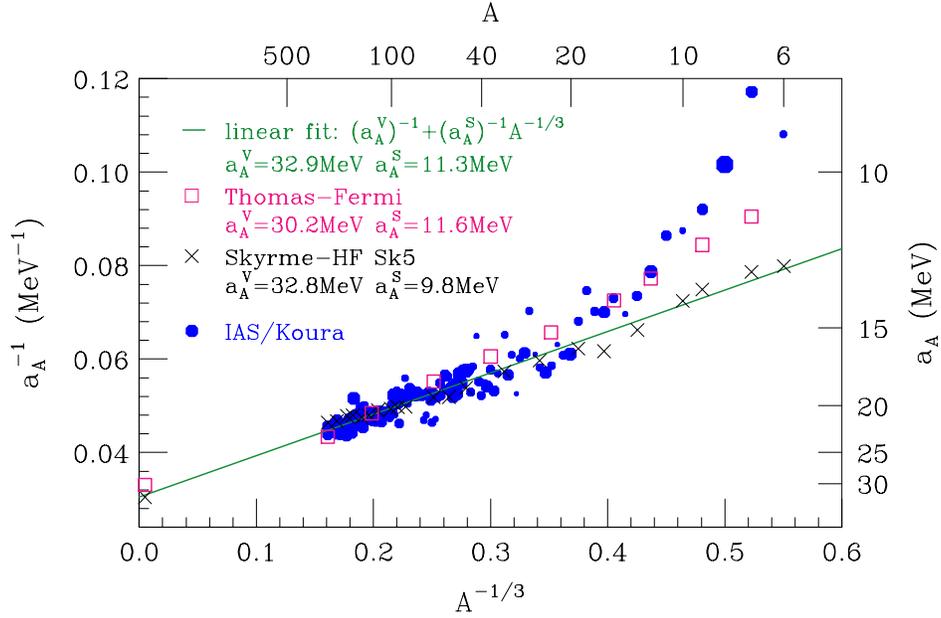}
\caption{Inverse of asymmetry coefficient, in the left scale, as a function of inverse cube root of mass number, in the bottom scale.  For convenience, top and right scale show, additionally, the mass number and the coefficient before inversion.  The filled circles represent coefficients extracted from IAS excitation energies with the microscopic corrections applied.  The line shows a fit to the IAS data for $A > 20$.  The crosses and squares represent, respectively, the coefficients from spherical Sk5 SHF calculations and coefficients from a Thomas-Fermi model that best describes the data.}
\label{fig:pa3}
\end{figure}

We use the tabulated energies of isobaric analog states \cite{Antony:1997} and microscopic corrections to energies by Koura \etal \cite{Koura:2005} to deduce symmetry coefficients for individual nuclear masses.  Those corrections include deformation effects.  Deduced coefficient values range from $a_a \sim 9 \, \text{MeV}$ for light $A < 10$ nuclei to $a_a \sim 22.5 \, \text{MeV}$ for $A > 200$.  Figure~\ref{fig:pa3} shows inverse coefficient values plotted against inverse cube root of mass number.  For $A > 20$, the value systematic is approximately linear, as expected from Eq.~(\ref{eq:AaaA}).  A fit with Eq.~(\ref{eq:AaaA}), produces coefficient values of $a_a^V = 32.9 \, \text{MeV}$ and $a_a^S = 11.3 \, \text{MeV}$.  Similar coefficient values are obtained when trying to describe the $a_a(A)$ data in terms of the Thomas-Fermi theory \cite{Danielewicz:2003dd,Danielewicz:2004et} or spherical SHF \cite{Reinhard:1991}.

\begin{figure}
  \includegraphics[height=.28\textheight]{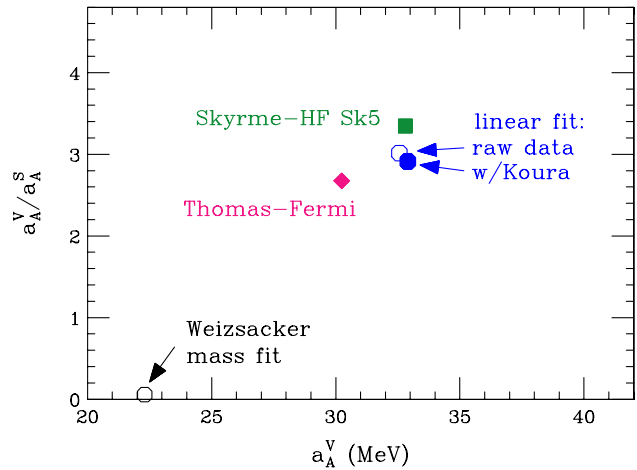}
\caption{Symmetry-energy parameter values in the plane of $a_a^V/a_a^S$ and $a_a^V$, from comparing theoretical expectations to data.}
\label{fig:rabd}
\end{figure}

The parameters arrived at in different ways are further summarized in Fig.~\ref{fig:rabd}.  The Thomas-Fermi results might be less reliable than other, because model yields an evolution of density profile in the surface region with mass, which is not fully supported by electron scattering data.  Overall, the IAS data suggest parameter values of $a_a^V = (31.5-33.5) \, \text{MeV}$ for the volume coefficient and $a_a^S = (9-12) \, \text{MeV}$.  Using the correlations between $a_a^V$ and $a_a^S$ coefficient values and $L$ and $K_\text{sym}$, such as in Fig.~\ref{fig:lavs}, within the Skyrme model, we further arrive at values of $L \sim 95 \, \text{MeV}$ and $K_\text{sym} \sim 25 \, \text{MeV}$.  These represent a nearly linear $S(\rho)$ at $\rho \lesssim \rho_0$.

Further efforts in this investigation strive to understand better the Coulomb effects for finite nuclei \cite{Danielewicz:2004et} and to exploit skin sizes as well as details in $\rho_p (r) $ for further constraining $S(\rho)$ at $\rho < \rho_0$.


\begin{theacknowledgments}
This work has been supported by the U.S.\ National Science Foundation under Grant Nos.\ PHY-0245009, PHY-0555893 and PHY-0606007.
\end{theacknowledgments}



\bibliographystyle{aipproc}   

\bibliography{cocu07}

\end{document}